# Statistical analysis of two arm randomized pre-post design with one post-treatment measurement

Fei Wan,[1] PhD

[1]Division of Public Health Sciences, Department of Surgery

Washington University School of Medicine

Saint Louis, MO, USA

**Corresponding Author:** Fei Wan

Division of Public Health Sciences, Department of Surgery

Washington University School of Medicine

Campus Box 8100

660 S. Euclid Ave, St. Louis, MO, USA

Tel: +1-314-362-9647; Fax: +1-314-747-3935

E-mail: wan.fei@wustl.edu

abstract
**Abstract**

*Introduction*: Randomized pre-post designs, with outcomes measured at baseline and follow-ups, have been commonly used to compare the clinical effectiveness of two competing treatments. There are vast, but often conflicting, amount of information in current literature about the best analytic methods for pre-post design. It is challenging for applied researchers to make an informed choice.

*Methods*: We discuss six methods commonly used in literature: one way analysis of variance ("*ANOVA"*), analysis of covariance main effect and interaction models on post-treatment measurement ("*ANCOVA* I" and "*ANCOVA* II"), *ANOVA* on change score between baseline and post-treatment measurements, repeated measures ("*RM"*) and constrained repeated measures ("*cRM"*) models on baseline and post-treatment measurements as joint outcomes. We review a number of study endpoints in pre-post designs and identify the difference in post-treatment measurement as the common treatment effect that all six methods target. We delineate the underlying differences and links between these competing methods in homogeneous and heterogeneous study population.

*Conclusion*: We demonstrate that *ANCOVA* and *cRM* outperform other alternatives because their treatment effect estimators have the smallest variances. *cRM* has comparable performance to *ANCOVA* I main effect model in homogeneous scenario and to *ANCOVA* II interaction model in heterogeneous scenario. In spite of that, *ANCOVA* has several advantages over *cRM,* including treating baseline measurement as covariate because it is not an outcome by definition, the convenience of incorporating other baseline variables and handling complex heteroscedasticity patterns in a linear regression framework.

# 1. Introduction

Two arm parallel randomized design has been widely used to compare the clinical effectiveness of competing treatments in improving patients' health outcomes. In these trials, continuous outcomes of interest are routinely measured at baseline and one follow up time point after treatment. Common statistical methods used in analyzing pre-post designs include: one way analysis of variance model (***ANOVA***),[1] analysis of covariance model adjusting for the baseline measurement (***ANCOVA* I**) [2-5] and ANCOVA including the baseline measurement by treatment interaction (***ANCOVA* II**) on post-treatment measurements[2,3], ***ANOVA*** on change score from the baseline (***ANOVA-Change***),[6,7] repeated measures models (***RM***) and constrained repeated measures models (***cRM***) on the baseline and post-treatment measurements as joint outcomes[8-10]. Despite of the simplicity and wide application of pre-post designs, which method is the best analytic approach has been a debated topic and many methodological studies have been performed to compare different statistical methods for past decades. However, it is challenging for applied researchers to evaluate this vast, but often conflicting, amount of information in current literature and make an informed choice.

The primary purpose of designing a pre-post randomized study is to answer the scientific question of interest: is treatment ***A*** more effective than treatment ***B***? To assess the difference in the treatment effectiveness between two treatments, we need to select a study endpoint and quantify a treatment effect. Common study endpoints include post treatment score, change score from baseline measurement to post treatment measurement, percent change from baseline, and rate of change from baseline. The difference between two arms on selected study endpoint is called treatment effect. Few studies examine the links between these different metrics of treatment effect. These underlying connections are critical in understanding the equivalence among some statistical methods that may appear to be very different at the first sight[2]. We need to be certain about the type of treatment effect each method targets and select the one that yields an unbiased and the most efficient estimator of the treatment effect of our interest.

In this study we aim to review six commonly used methods (**ANOVA, ANCOVA I, ANCOVA II, ANOVA-Change, RM, cRM**) from a practical standpoint, and we focus on delineating the differences and underlying connections between these methods. In section (2), we provide notations and assumptions for a typical pre-post design, define homogeneous and heterogeneous study population, and discuss common study endpoints and associated treatment effects. In section (3), we first outline these competing models using the same set of population mean, variance, and covariance parameters, and then assess differences and links between them in homogeneous and heterogeneous scenarios; In section (4), we present three simulated weight loss trial data examples (homogeneous data, heterogeneous data with balanced design, heterogeneous data with unbalanced design) to exemplify the differences and links between different statistical methods. In the last section, we discuss the results and give recommendation on the best analytical approach in pre-post designs.

## 2. A hypothetical weight loss trial and metrics of treatment effect

*2.1 Notations*

In a hypothetical two arm parallel weight loss trial comparing the effect of a new drug ("treatment") and a placebo ("control") in reducing participants' weights, we use $Y_{ijt}$ to denote weight of the $i$th patient ($i = 1,2,3, \ldots n_j$) in the treatment arm $j$ ($j = 0,1$) at equally spaced time point $t$ ($t = t_0, t_1$). $n_0$ and $n_1$ are the number of subjects in the control and treatment arms.

We denote the mean baseline weights for the treatment and control arms by $\mu_{1t_0}$ and $\mu_{0t_0}$, respectively. It is reasonable to assume $\mu_{1t_0} = \mu_{0t_0}$ under randomization and we let $\mu_{t_0}$ denote the common baseline weight mean. Mean weights at the follow-up time point $t_1$ in the treatment and control arms are denoted by $\mu_{1t_1}$ and $\mu_{0t_1}$, respectively (Figure 1). We define homogeneous and heterogeneous study populations as follows:

i) ***The homogeneous scenario:*** every subject, in either the treatment or control arm, has the same pattern of variance and covariance structure for their baseline and follow-up weights.

$$\Sigma = \begin{bmatrix} \sigma_0^2 & \rho\sigma_0\sigma_1 \\ \rho\sigma_0\sigma_1 & \sigma_1^2 \end{bmatrix},$$

where $\sigma_0^2$ and $\sigma_1^2$ are the variances of weight measurements at baseline and follow-up, $\rho$ is the correlation coefficient between the baseline and follow-up weights.

ii) ***The heterogeneous scenario:*** variance and covariance structures of the baseline and follow-up weights differ between the treatment and control arms. The variance and covariance matrix for the control arm is

$$\Sigma_0 = \begin{bmatrix} \sigma_0^2 & \rho_0\sigma_0\sigma_{01} \\ \rho_0\sigma_0\sigma_{01} & \sigma_{01}^2 \end{bmatrix},$$

The variance and covariance matrix for the treatment arm is

$$\Sigma_1 = \begin{bmatrix} \sigma_0^2 & \rho_1\sigma_0\sigma_{11} \\ \rho_1\sigma_0\sigma_{11} & \sigma_{11}^2 \end{bmatrix},$$

where $\sigma_0^2$ is the common variance of weights at baseline for both control and treatment arms. Both arms have the same baseline variances because of randomization at baseline. $\sigma_{01}^2$ and $\sigma_{11}^2$ are variances of weight measurements at follow-up in the control and treatment arms. $\rho_0$ and $\rho_1$ are the correlation coefficients between weight measurements at baseline and follow-up in the control and treatment arms, respectively. In practice, the correlation between the pre- and post-treatment measurements are usually stronger in the control arm than in the treatment arm.

Variability of post-treatment measurement tends to be larger in the treatment arm than in the control arm because participants may respond to the treatment more differently. i.e., $\rho_0 > \rho_1$ and $\sigma_{11}^2 > \sigma_{01}^2$.

*2.2 Metrics of treatment effect*

We discuss the following three commonly reported metrics of treatment effect in pre-post trials:

i)     The primary endpoint is weight at the follow-up time point $t_1$. The difference in mean weights at $t_1$ between two arms is the treatment effect:

$$\tau = \mu_{1t_1} - \mu_{0t_1}$$

e.g, if $\tau = -10$, we can interpret the results as "at the end of the trial, the mean weight was 10 pounds lower in the treatment group than in the control group."

ii)     The primary endpoint is change score calculated by subtracting the follow up weight from the baseline weight $\Delta_{ij} = Y_{ijt_1} - Y_{ijt_0}$. The mean difference in change scores between two arms is the treatment effect. Formally, we have:

$$\tilde{\tau} = (\mu_{1t_1} - \mu_{1t_0}) - (\mu_{0t_1} - \mu_{0t_0})$$

e.g. if $\tilde{\tau} = -10$, this difference is normally interpreted as "weight **reductions** were 10 pounds greater on treatment than control.". Since we have $\mu_{0t_1} = \mu_{0t_0}$ due to randomization, it follows directly $\tilde{\tau} = \tau$. Furthermore, when we code "0" for the baseline time point ($t_0 = 0$) and "1" for the follow-up time point ($t_1 = 1$), mean change score for each arm can also be interpreted as mean change rate per unit time for each arm, represented by slopes in Figure (1). Thus, difference in slopes, denoted by $\tilde{\tilde{\tau}} = \alpha_1 - \alpha_0$, is also equivalent to $\tau$. As we will discuss in details in section (3), ***ANOVA*** and ***ANCOVA*** target $\tau$, ***ANOVA-CHANGE*** targets $\tilde{\tau}$, and ***RM*** targets $\tilde{\tilde{\tau}}$. We can compare these seemingly very different statistical methods in a meaningful way because of the equivalence between $\tau$, $\tilde{\tau}$, and $\tilde{\tilde{\tau}}$ in randomized pre-post designs.

iii) The primary endpoint is the percent change from baseline weight, denoted by $\varphi_{ij} = \frac{(Y_{ijt_1} - Y_{ijt_0})}{Y_{ijt_0}}$. Treatment effect is the mean difference in the percentage change between two arms and formally defined as follows:

$$\tau^* = \bar{\varphi}_1 - \bar{\varphi}_0,$$

where $\bar{\varphi}_1$ and $\bar{\varphi}_0$ are the mean percentage change in the treatment and control arms. Although percent change is popular among clinical researchers, this metric has several drawbacks[11-13]. First, percent change is a function of ratio $\frac{Y_{ijt_1}}{Y_{ijt_0}}$. The distribution of percent change is highly skewed. Analyzing it with normal-theory based statistical methods is not justified and non-parametric statistical methods are generally less powerful; Second, percent change is not a symmetric and additive measure. For example, mean weight of adults over 20 in US is 197.8 pound for men and 170.5 pound for women. The mean difference is 27.3 pound between men and women. Men weight 100×((197.8-170.5)/170.5)=16% more than women, whereas women weight 100×((197.8-170.5)/197.8)=13.8% less than men. The percentage differences could be different depending on which sex is used as devisor. If a participant's weight increases by 10% in first 6 months and fall by 10% for the next 6 months, the two percentage changes do not cancel out. The participant's weight at the end would be only 99% of the participant's starting weight.

## 3. Statistical methods

In sections (3.1) and (3.2), we focus on six methods aiming to estimate $\tau$. We describe each statistical model using the same set of population mean, variance, and covariance parameters defined in section (2) for homogeneous and heterogeneous scenarios, separately. We present the closed-form expressions of each point estimator of treatment effect and its variance. It often goes unnoticed in practice that different statistical methods target different types of variances (i.e., conditional vs. unconditional variances) for the corresponding treatment effect estimators. Baseline measurement is random because we rarely can fix the baseline characteristics of participants when we enroll them into randomized trials. Thus, unconditional variance and corresponding unconditional inference is of greater interest. However, covariate adjusted OLS model-based variances for ANCOVA are conditional because OLS assumes adjusted covariates are fixed. We will discuss in details whether OLS model-based conditional inference (i.e., test statistics and *p*-values from standard statistical software) is still valid for unconditional hypothesis testing and the potential fixes that we can use to draw valid unconditional inference if the usual OLS model-based inference is biased.

*3.1 When the study population is homogeneous*

**Method 1:** *ANOVA modeling post treatment measure ("ANOVA-Post").* We model weight at follow-up $Y_{ijt_1}$ using the binary treatment indicator $G_{ij}$ (1 if in the treatment arm; 0 if in the control arm) as follows:

$$Y_{ijt_1} = \beta_0^{(1)} + \beta_1^{(1)} G_{ij} + e_{ij}^{(1)}, i = 1,2,\ldots,n_j; j = 0,1; \quad (1)$$

$$e_{ij}^{(1)} \sim N(0, \sigma_1^2),$$

where $\beta_0^{(1)} = \mu_{0t_1}$, $\beta_1^{(1)} = \mu_{1t_1} - \mu_{0t_1} = \tau$, and $e_{ij}^{(1)}$ is independently and identically distributed (i.i.d) random error. $\beta_1^{(1)}$ represents the treatment effect. Model (1) is homoscedastic with a constant residual variance $\sigma_1^2$.

We fit an ordinary least squares (OLS) regression to estimate the coefficients and standard errors for model (1). The closed-form expressions of the OLS estimator $\hat{\beta}_{1,ols}^{(1)}$ and its variance, denoted by $var(\hat{\beta}_{1,ols}^{(1)})$, are presented in Table 1. $\hat{\beta}_{1,ols}^{(1)}$ is estimated by the sample group mean difference in post-treatment weight between two arms. $\hat{\beta}_{1,ols}^{(1)}$ is unbiased for $\tau$. The OLS model-based variance of $\hat{\beta}_{1,ols}^{(1)}$ assuming known $\sigma_1^2$ is expressed as follows:

$$var_{ols}(\hat{\beta}_{1,ols}^{(1)}) = \frac{\sigma_1^2}{\sum_{j=0}^{1}\sum_{i=1}^{n_j}(G_{ij}-G_{..})^2},$$

where $G_{..} = \frac{\sum_{j=0}^{1}\sum_{i=1}^{n_j} G_{ij}}{n_0+n_1} = \frac{n_1}{n_0+n_1}$. $\sigma_1^2$ is estimated by

$$\hat{\sigma}_1^2 = \frac{\sum_{j=0}^{1}\sum_{i=1}^{n_j}\left(y_{ijt_1}-\hat{y}_{ijt_1}^{(1)}\right)^2}{(n_0+n_1-2)},$$

where $\hat{y}_{ijt_1}^{(1)} = \hat{\beta}_{0,ols}^{(1)} + \hat{\beta}_{1,ols}^{(1)} G_{ij}$ is the predicted value from fitted model (1). We let $\widehat{var}_{ols}(\hat{\beta}_{1,ols}^{(1)})$ denote the OLS model-based variance estimator with the estimator $\hat{\sigma}_1^2$, which is output by standard statistical software (Table 1). Since $\sum_{j=0}^{1}\sum_{i=1}^{n_j}(G_{ij}-G_{..})^2 = \frac{n_0 n_1}{n_0+n_1}$, it follows that $var_{ols}(\hat{\beta}_{1,ols}^{(1)}) = var(\hat{\beta}_{1,ols}^{(1)})$. It is well established that $\widehat{var}_{ols}(\hat{\beta}_{1,ols}^{(1)})$ is unbiased for $var_{ols}(\hat{\beta}_{1,ols}^{(1)})$ in a homoscedastic linear model. Thus, $\widehat{var}_{ols}(\hat{\beta}_{1,ols}^{(1)})$ is unbiased for $var(\hat{\beta}_{1,ols}^{(1)})$. The usual OLS model-based inference (i.e., test statistics $t = \frac{\hat{\beta}_{1,ols}^{(1)}}{\sqrt{\widehat{var}_{ols}(\hat{\beta}_{1,ols}^{(1)})}}$ and the associated p-value) is valid for testing $H_o: \tau = 0$ unconditionally.

**Method 2:** *ANCOVA modeling post treatment measure ("ANCOVA I"):* We model the follow-up weight $Y_{ijt_1}$ using the binary treatment indicator $G_{ij}$ and the baseline weight $Y_{ijt_0}$.

$$Y_{ijt_1} = \beta_0^{(2)} + \beta_1^{(2)} G_{ij} + \beta_2^{(2)} Y_{ijt_0} + e_{ij}^{(2)}, i=1,2,\ldots,n_j; j=0,1; \quad (2)$$

$$e_{ij}^{(2)} \sim N(0, \sigma_{\epsilon^{(2)}}^2) \text{ and } \sigma_{\epsilon^{(2)}}^2 = (1-\rho^2)\sigma_1^2$$

, where $\beta_0^{(2)} = \mu_{0t_1} - \rho\frac{\sigma_1}{\sigma_0}\mu_{t_0}$, $\beta_1^{(2)} = \tau$, $\beta_2^{(2)} = \rho\frac{\sigma_1}{\sigma_0}$, and $e_{ij}^{(2)}$ is i.i.d random error. $\beta_1^{(2)}$ measures the treatment effect $\tau$ and $\beta_2^{(2)}$ represents the slope of the pre-post association between $Y_{ijt_1}$ and $Y_{ijt_0}$. OLS assumes every subject has a common residual variance $\sigma_{\epsilon^{(2)}}^2$. Model (2) implicitly assumes that two arms share the common baseline mean $\mu_{t_0}^2$.

We fit an OLS regression to estimate the regression coefficients and standard errors for model (2). The OLS estimator $\hat{\beta}_{1,ols}^{(2)}$ is derived as the sample mean difference in post-treatment weight adjusting for the difference in sample mean baseline weight between two arms. The mean difference in baseline weight between two arms can be seen as chance imbalance in a randomized trial. $\hat{\beta}_{1,ols}^{(2)}$ is unbiased for $\tau$ both conditional on $Y_{ijt_0}$ and unconditionally. The formulas of $\hat{\beta}_{1,ols}^{(2)}$ and its "unconditional" variance $var(\hat{\beta}_{1,ols}^{(2)})$ are listed in Table 1. However, OLS assumes adjusted baseline weight $Y_{ijt_0}$ is fixed. OLS targets the conditional variance of $\hat{\beta}_{1,ols}^{(2)}$, denoted by $var(\hat{\beta}_{1,ols}^{(2)}|Y_{ijt_0})$, instead of $var(\hat{\beta}_{1,ols}^{(2)})$. The formula of $var(\hat{\beta}_{1,ols}^{(2)}|Y_{ijt_0})$ with known common residual variance $\sigma_{\epsilon^{(2)}}^2$ is presented in Table 1. Since $\sigma_{\epsilon^{(2)}}^2$ is generally unknown, it is estimated by the following sample residual variance:

$$\hat{\sigma}_{e_{ij}^{(2)}}^2 = \frac{\sum_{j=0}^{1}\sum_{i=1}^{n_j}\left(y_{ijt_1} - \hat{y}_{ijt_1}^{(2)}\right)^2}{(n_0 + n_1 - 3)}$$

, where $\hat{y}_{ijt_1}^{(2)} = \hat{\beta}_{0,ols}^{(2)} + \hat{\beta}_{1,ols}^{(2)}G_{ij} + \hat{\beta}_{2,ols}^{(2)}Y_{ijt_0}$, the predicted value from fitted model (2). We let $\widehat{var}_{ols}(\hat{\beta}_{1,ols}^{(2)}|Y_{ijt_0})$ denote the OLS model-based variance estimator with the estimator $\hat{\sigma}_{\epsilon^{(2)}}^2$. $\widehat{var}_{ols}(\hat{\beta}_{1,ols}^{(2)}|Y_{ijt_0})$ is output by standard statistical software (e.g. "proc reg" in SAS). Its formula is presented in Table 1.

Because we want to generalize our conclusions to study populations where values of $Y_{ijt_0}$ can be different from the values in our current sample, we may wonder whether significance tests based on the conditional variance assuming $Y_{ijt_0}$ is fixed (e.g., $t = \frac{\hat{\beta}^{(2)}_{1,ols}}{\sqrt{\widehat{var}_{ols}(\hat{\beta}^{(2)}_{1,ols}|Y_{ijt_0})}}$) is comparable to unconditional inference (e.g., $t = \frac{\hat{\beta}^{(2)}_{1,ols}}{\sqrt{var(\hat{\beta}^{(2)}_{1,ols})}}$), in which $Y_{ijt_0}$ is treated as random variable, for testing $H_o: \tau = 0$. To establish this equivalence, we need to show: i) $\widehat{var}_{ols}(\hat{\beta}^{(2)}_{1,ols}|Y_{ijt_0})$ is unbiased for $var(\hat{\beta}^{(2)}_{1,ols}|Y_{ijt_0})$; ii) $var(\hat{\beta}^{(2)}_{1,ols}|Y_{ijt_0})$ is unbiased for $var(\hat{\beta}^{(2)}_{1,ols})$. The first part is well established in a homoscedastic linear model. The second part holds because $var(\hat{\beta}^{(2)}_{1,ols}) = E(var(\hat{\beta}^{(2)}_{1,ols}|Y_{ijt_0}))$. That is, unconditional variance of $\hat{\beta}^{(2)}_{1,ols}$ is the average of its conditional variance over the distribution of baseline weight measurement. Therefore, the usual model-based standard errors and associated *p*-values are valid for unconditional inference[2,4].

**Method 3: *Repeated measures model ("RM")*:** Different from modeling baseline weight as a covariate in model (2), we model the baseline and follow up weights ($Y_{ijt_0}$, $Y_{ijt_1}$) jointly using the binary treatment indicator $G_{ij}$, the binary time factor $T_{ij}$, the time by treatment interaction $G_{ij} \times T_{ij}$ in the following ***RM*** model:

$$Y_{ijt} = \gamma_0^{(3)} + \gamma_1^{(3)} G_{ij} + \gamma_2^{(3)} T_{ij} + \gamma_3^{(3)} G_{ij} \times T_{ij} + e_{ijt}^{(3)}, i = 1,2,\ldots,n_j; j = 0,1; t = t_0, t_1 \quad (3)$$

$$\begin{pmatrix} e_{ijt_0}^{(3)} \\ e_{ijt_1}^{(3)} \end{pmatrix} \sim N\left(\begin{bmatrix} 0 \\ 0 \end{bmatrix}, \Sigma\right).$$

When $t_0 = 0$ and $t_1 = 1$, $\gamma_0^{(3)} = \mu_{0t_0}$, $\gamma_1^{(3)} = \mu_{1t_0} - \mu_{0t_0}$, $\gamma_2^{(3)} = \mu_{0t_1} - \mu_{0t_0}$, and $\gamma_3^{(3)} = (\mu_{1t_1} - \mu_{1t_0}) - (\mu_{0t_1} - \mu_{0t_0})$. $\gamma_0^{(3)}$ represents the mean baseline weight for the control arm, $\gamma_1^{(3)}$ represents the difference in mean baseline weights between the treatment and control arms, $\gamma_2^{(3)}$ represents the mean change from the baseline weight in the control arm, and $\gamma_3^{(3)}$ is generally interpreted as the difference in the mean change from baseline weight in a unit time interval between the treatment and control arms ("difference in difference"), also known as the difference in slopes. Under the randomization, we have $\mu_{1t_0} = \mu_{0t_0}$ and it follows that $\gamma_1^{(3)} = 0$ and $\gamma_3^{(3)} = \mu_{1t_1} - \mu_{1t_1} = \tau$. Thus, testing $H_o: \gamma_3^{(3)} = 0$ is equivalent to testing $H_o: \tau = 0$.

We fit a generalized least squares (GLS) model with correlated outcomes to estimate the coefficients and standard errors for model (3). The closed-form expressions of the GLS estimator of treatment effect $\hat{\gamma}_{3,gls}^{(3)}$ and its variance $var(\hat{\gamma}_{3,gls}^{(3)})$ given known variance and covariance parameters are presented in Table 1. $\hat{\gamma}_{3,gls}^{(3)}$ is estimated by the sample mean difference in weight change between two arms and is unbiased for $\tau$ asymptotically (i.e. in large sample). Variance and covariance parameters are generally unknown and need to be estimated using restricted maximum likelihood (REML). Conventional maximal likelihood estimation (MLE) should be avoided. The REML variance estimator $\widehat{var}_{reml}(\hat{\gamma}_{3,gls}^{(3)})$ is derived by plugging the REML estimators of variance and covariance parameters (i.e., $\sigma_0^2, \sigma_1^2, \rho\sigma_0\sigma_1$) into the formula of $var(\hat{\gamma}_{3,gls}^{(3)})$. We use Kenward and Roger method[15] ("ddfm=kenwardroger" in SAS proc mixed procedure) to adjust for the potential finite sample bias in $\widehat{var}_{reml}(\hat{\gamma}_{3,gls}^{(3)})$ because of its failure to incorporate variabilities of the REML estimators of variance and covariance parameters. This adjustment involves inflating variance and covariance matrix and computing an adjusted approximation degrees of freedom.

**Method 4:** ***Constrained Repeated measures Model ("cRM"):*** By specifying $\gamma_1^{(3)}$ in the model, ***RM*** model (3) assumes mean baseline weight is different between two arms. Liang and Zeger[8] proposed the following *cRM* model by fixing $\gamma_1^{(3)} = 0$ to force the treatment and control arms to have the same intercept. Intuitively, *cRM* is more efficient than *RM* because *cRM* estimates one less parameter. Formally, we model baseline and follow-up weights $(Y_{ijt_0}, Y_{ijt_1})$ jointly using the binary factor $T_{ij}$, time by treatment interaction $G_{ij} \times T_{ij}$ in the following *cRM* model:

$$Y_{ijt} = \gamma_0^{(4)} + \gamma_2^{(4)} T_{ij} + \gamma_3^{(4)} G_{ij} \times T_{ij} + e_{ijt}^{(4)}, i = 1, 2, \ldots, n_j; j = 0, 1; t = t_0, t_1 \quad (4)$$

$$\begin{pmatrix} e_{ijt_0}^{(4)} \\ e_{ijt_1}^{(4)} \end{pmatrix} \sim N \left( \begin{bmatrix} 0 \\ 0 \end{bmatrix}, \Sigma \right),$$

where $\gamma_0^{(4)} = \mu_{t_0}, \gamma_2^{(4)} = \mu_{0t_1} - \mu_{0t_0}$, and $\gamma_3^{(4)} = \tau$. Interpretations of $\gamma_0^{(4)}$, $\gamma_2^{(4)}$, and $\gamma_3^{(4)}$ are the same as their counterparts in *RM*. The formulas of the GLS point estimator $\hat{\gamma}_{3,gls}^{(4)}$ and its variance $var(\hat{\gamma}_{3,gls}^{(4)})$ are listed in Table 1. $\hat{\gamma}_{3,gls}^{(4)}$ is unbiased for $\tau$ asymptotically. Empirical or model-based variance estimate for $var(\hat{\gamma}_{3,gls}^{(4)})$ is derived in the same way as regular *RM* model using REML.

**Method 5: *ANOVA with change score ("ANOVA-Change")*:** We model change score $\Delta_{ij} = Y_{ijt_1} - Y_{ijt_0}$ using the binary treatment indicator $G_{ij}$ as follows:

$$\Delta_{ij} = \beta_0^{(5)} + \beta_1^{(5)} G_{ij} + e_{ij}^{(5)}, i = 1, 2, \ldots, n_j; j = 0, 1; \quad (5)$$

$$e_{ij}^{(5)} \sim N(0, \sigma_{\epsilon^{(5)}}^2) \text{ and } \sigma_{\epsilon^{(5)}}^2 = \sigma_1^2 + \sigma_0^2 - 2\rho\sigma_0\sigma_1,$$

where $\beta_0^{(5)} = \mu_{0t_1} - \mu_{0t_0}$, $\beta_1^{(5)} = (\mu_{1t_1} - \mu_{1t_0}) - (\mu_{0t_1} - \mu_{0t_0})$, and $e_{ij}^{(3)}$ is i.i.d random error. $\beta_0^{(5)}$ measures the mean difference score in the control arm. $\beta_1^{(5)}$ measures the treatment effect $\tilde{\tau}$. Since $\mu_{1t_0} = \mu_{0t_0}$ due to randomization at baseline, $\beta_1^{(5)}$ is reduced to $\tau$. The closed-form expressions of $\hat{\beta}_{1,ols}^{(5)}$ and $var(\hat{\beta}_{1,ols}^{(5)})$ are listed in Table 1. $\hat{\beta}_{1,ols}^{(5)}$ is derived as the sample mean difference in weight change between two arms ("difference in difference") and is unbiased for $\tau$. The OLS model-based variance of $\hat{\beta}_{1,ols}^{(5)}$ assuming known $\sigma_{\epsilon^{(5)}}^2$ is

$$var_{ols}(\hat{\beta}_{1,ols}^{(5)}) = \frac{\sigma_{\epsilon^{(5)}}^2}{\sum_{j=0}^{1}\sum_{i=1}^{n_j}(G_{ij} - G_{..})^2},$$

where $G_{..} = \frac{\sum_{j=0}^{1}\sum_{i=1}^{n_j} G_{ij}}{n_0 + n_1} = \frac{n_1}{n_0 + n_1}$. $\sigma_{\epsilon^{(5)}}^2$ is estimated by

$$\hat{\sigma}_{\epsilon^{(5)}}^2 = \frac{\sum_{j=0}^{1}\sum_{i=1}^{n_j}\left(\Delta_{ij} - \hat{\Delta}_{ij}^{(5)}\right)^2}{(n_0 + n_1 - 2)},$$

where $\hat{\Delta}_{ij}^{(5)}$ is the fitted value from OLS model (5). We let $\widehat{var}_{ols}(\hat{\beta}_{1,ols}^{(5)})$ denote the OLS model-based variance estimator with the estimator $\hat{\sigma}_{\epsilon^{(5)}}^2$, which is output by standard statistical software (Table 1). Since $\sum_{j=0}^{1}\sum_{i=1}^{n_j}(G_{ij} - G_{..})^2 = \frac{n_0 n_1}{n_0 + n_1}$, it follows that $var_{ols}(\hat{\beta}_{1,ols}^{(5)}) = var(\hat{\beta}_{1,ols}^{(5)})$. It is well established that $\widehat{var}_{ols}(\hat{\beta}_{1,ols}^{(5)})$ is unbiased for $var_{ols}(\hat{\beta}_{1,ols}^{(5)})$, and thus for $var(\hat{\beta}_{1,ols}^{(5)})$. The usual OLS model-based inference is valid for unconditional hypothesis testing.

*3.2 When the study population is heterogeneous*

**Method 6: *ANCOVA* II:** Different variance and covariance structures in the treatment and control arms suggest a baseline measurement by treatment interaction term in ANCOVA[2,3,9,10]. To estimate $\tau$ using an interaction model, we first compute the mean centered baseline weight $\tilde{Y}_{ijt_0}$ by subtracting overall mean baseline weight from individual baseline weight. i.e., $\tilde{Y}_{ijt_0} = Y_{ijt_0} - \mu_{t_0}$. We then model the post-treatment weight $Y_{ijt_1}$ using the binary treatment indicator $G_{ij}$, the mean centered baseline weight $\tilde{Y}_{ijt_0}$, and the baseline by treatment interaction $G_{ij} \times \tilde{Y}_{ijt_0}$ as follows:

$$Y_{ijt_1} = \beta_0^{(6)} + \beta_1^{(6)} G_{ij} + \beta_2^{(6)} \tilde{Y}_{ijt_0} + \beta_3^{(6)} G_{ij} \times \tilde{Y}_{ijt_0} + e_{ij}^{(6)}, i = 1,2,\ldots,n_j; j = 0,1; \quad (6)$$

$$e_{i0}^{(6)} \sim N(0, \sigma_{\epsilon_0^{(6)}}^2) \text{ and } \sigma_{\epsilon_0^{(6)}}^2 = (1-\rho_0^2)\sigma_{01}^2$$

$$e_{i1}^{(6)} \sim N(0, \sigma_{\epsilon_1^{(6)}}^2) \text{ and } \sigma_{\epsilon_1^{(6)}}^2 = (1-\rho_1^2)\sigma_{11}^2$$

, where $\beta_0^{(6)} = \mu_{0t_1}$, $\beta_1^{(6)} = \tau$, $\beta_2^{(6)} = \rho_0 \frac{\sigma_{0t_0}}{\sigma_0}$, and $\beta_3^{(6)} = \rho_1 \frac{\sigma_{1t_1}}{\sigma_0} - \rho_0 \frac{\sigma_{0t_0}}{\sigma_0}$. $e_{i0}^{(6)}$ and $e_{i1}^{(6)}$ are i.i.d random errors in the control and treatment arms. $\beta_1^{(6)}$ is the parameter associated with the treatment effect. $\beta_2^{(6)}$ is the regression slope of baseline weight in the control arm. $\beta_3^{(6)}$ measures the difference in regression slopes of baseline weight between the treatment and control arms. Model (6) is heteroscedastic because error terms in the treatment and control arms have different residual variances.

The OLS estimator $\hat{\beta}_{1,ols}^{(6)}$ is the adjusted mean difference in post-treatment weights controlling for a weighted mean difference of baseline weights between two arms with unequal weights for treatment and control arms (i.e., $\hat{\beta}_{2,ols}^{(6)} + \hat{\beta}_{3,ols}^{(6)}$ for the treatment group, and $\hat{\beta}_{2,ols}^{(6)}$ for the control group) (Table 2). $\hat{\beta}_{1,ols}^{(6)}$ is unbiased for $\tau$. The conditional variance of $\hat{\beta}_{1,ols}^{(6)}$, denoted by $var(\hat{\beta}_{1,ols}^{(6)}|\tilde{Y}_{ijt_0})$, incorporates two residual variances $\sigma_{\epsilon_0^{(6)}}^2$ and $\sigma_{\epsilon_1^{(6)}}^2$ (Table 2). Standard statistical software such as SAS does not output $var(\hat{\beta}_{1,ols}^{(6)}|\tilde{Y}_{ijt_0})$ because OLS assumes a common residual variance $\sigma_{\epsilon^{(6)}}^2$, which is the following weighted average of $\sigma_{\epsilon_0^{(6)}}^2$ and $\sigma_{\epsilon_1^{(6)}}^2$:

$$\sigma_{\epsilon^{(6)}}^2 = \frac{n_0}{n_0+n_1} \sigma_{\epsilon_0^{(6)}}^2 + \frac{n_1}{n_0+n_1} \sigma_{\epsilon_1^{(6)}}^2.$$

We let $var_{ols}(\hat{\beta}_{1,ols}^{(6)}|\tilde{Y}_{ijt_0})$ denote the OLS model-based conditional variance of $\hat{\beta}_{1,ols}^{(6)}$ with known $\sigma_{\epsilon^{(6)}}^2$ (Table 2). Since $\sigma_{\epsilon^{(6)}}^2$ is generally unknown, $\sigma_{\epsilon^{(6)}}^2$ is estimated by

$$\hat{\sigma}^2_{\epsilon(6)} = \frac{\sum_{j=0}^{1}\sum_{i=1}^{n_j}(y_{ijt_1}-\hat{y}_{ijt_1})^2}{(n_0+n_1-4)},$$

where $\hat{y}_{ijt_1}$ is the predicted value of $y_{ijt_1}$. We let $\widehat{var}_{ols}(\hat{\beta}^{(6)}_{1,ols}|\tilde{Y}_{ijt_0})$ denote the OLS model-based variance estimator of $\hat{\beta}^{(6)}_{1,ols}$ with the estimator $\hat{\sigma}^2_{\epsilon(6)}$ and known constant $\mu_{t_0}$ (Table 2). $\widehat{var}_{ols}(\hat{\beta}^{(6)}_{1,ols}|\tilde{Y}_{ijt_0})$ is output by standard statistical software (e.g., "proc reg" in SAS). To assess the validity of model-based standard errors and $p$-values from regular *ANCOVA* II model for unconditional inference, we need to examine: i) whether $\widehat{var}_{ols}(\hat{\beta}^{(6)}_{1,ols}|\tilde{Y}_{ijt_0})$ is unbiased for $var(\hat{\beta}^{(6)}_{1,ols}|\tilde{Y}_{ijt_0})$; ii) whether $var(\hat{\beta}^{(6)}_{1,ols}|\tilde{Y}_{ijt_0})$ is unbiased for $var(\hat{\beta}^{(6)}_{1,ols})$.

First, $\widehat{var}_{ols}(\hat{\beta}^{(6)}_{1,ols}|\tilde{Y}_{ijt_0})$ is unbiased for $var_{ols}(\hat{\beta}^{(6)}_{1,ols}|\tilde{Y}_{ijt_0})$. However, the unbiasedness of $\widehat{var}_{ols}(\hat{\beta}^{(6)}_{1,ols}|\tilde{Y}_{ijt_0})$ as an estimator of $var(\hat{\beta}^{(6)}_{1,ols}|\tilde{Y}_{ijt_0})$ depends on the relationship between $var_{ols}(\hat{\beta}^{(6)}_{1,ols}|\tilde{Y}_{ijt_0})$ and $var(\hat{\beta}^{(6)}_{1,ols}|\tilde{Y}_{ijt_0})$. It can be shown in balanced design ($n_0 = n_1$),

$$var_{ols}(\hat{\beta}^{(6)}_{1,ols}|\tilde{Y}_{ijt_0}) \approx var(\hat{\beta}^{(6)}_{1,ols}|\tilde{Y}_{ijt_0}).$$

Thus, $\widehat{var}_{ols}(\hat{\beta}^{(6)}_{1,ols}|\tilde{Y}_{ijt_0})$ is nearly unbiased for $var(\hat{\beta}^{(6)}_{1,ols}|\tilde{Y}_{ijt_0})$.[2] When the design is unbalanced ($n_0 \neq n_1$),

$$var_{ols}(\hat{\beta}^{(6)}_{1,ols}|\tilde{Y}_{ijt_0}) \neq var(\hat{\beta}^{(6)}_{1,ols}|\tilde{Y}_{ijt_0}).$$

Hence, $\widehat{var}_{ols}(\hat{\beta}^{(6)}_{1,ols}|\tilde{Y}_{ijt_0})$ is biased for $var(\hat{\beta}^{(6)}_{1,ols}|\tilde{Y}_{ijt_0})$. $\widehat{var}_{ols}(\hat{\beta}^{(6)}_{1,ols}|\tilde{Y}_{ijt_0})$ over-estimates $var(\hat{\beta}^{(6)}_{1,ols}|\tilde{Y}_{ijt_0})$ if the group with larger residual variance has larger sample size and the group with smaller residual variance has smaller sample size, and otherwise may underestimate $var(\hat{\beta}^{(6)}_{1,ols}|\tilde{Y}_{ijt_0})$ [2,3].

Second, common mean baseline weight $\mu_{t_0}$ is generally unknown. We need to estimate $\mu_{t_0}$ in $\tilde{Y}_{ijt_0}$ using overall sample mean $\hat{\mu}_{t_0} = \frac{\sum_{j=0}^{1} \sum_{i=1}^{n_j} Y_{ijt_0}}{n_0+n_1}$ but ANCOVA treats $\hat{\mu}_{t_0}$ as fixed and fails to capture this additional variability in conditional variances. As shown below, it turns out that $var(\hat{\beta}_{1,ols}^{(6)}|\tilde{Y}_{ijt_0})$ underestimates $var(\hat{\beta}_{1,ols}^{(6)})$ by a factor of $\beta_{3,ols}^{[6]2} var(\hat{\mu}_{t_0})^2$:

$$var(\hat{\beta}_{1,ols}^{(6)}) = E(var(\hat{\beta}_{1,ols}^{(6)}|\tilde{Y}_{ijt_0})) + \beta_{3,ols}^{[6]2} var(\hat{\mu}_{t_0}).$$

Thus, the OLS model-based conditional inference is biased for unconditional hypothesis testing because of heteroscedasticity and neglecting of sampling variability of $\hat{\mu}_{t_0}$. To fix these two problems, we can use the following adjusted heteroscedasticity-consistent (HC) variance estimator to replace $\widehat{var}_{ols}(\hat{\beta}_{1,ols}^{(6)}|\tilde{Y}_{ijt_0})$ for valid unconditional inference:

$$\widehat{var}_{aHC}(\hat{\beta}_{1,ols}^{(6)}|\tilde{Y}_{ijt_0}) = \widehat{var}_{HC}(\hat{\beta}_{1,ols}^{(6)}|\tilde{Y}_{ijt_0}) + \hat{\beta}_{3,ols}^{[6]2} \frac{\hat{\sigma}_0^2}{n_0+n_1},$$

where $\widehat{var}_{HC}(\hat{\beta}_{1,ols}^{(6)}|\tilde{Y}_{ijt_0})$ is a HC variance estimator for $var(\hat{\beta}_{1,ols}^{(6)}|\tilde{Y}_{ijt_0})$ [16] and can be output from standard software. HC variance estimators are consistent (i.e., unbiased in large sample). Among all available HC variance estimators, HC2 was shown to have the best performance in finite samples[2,3] (e.g. "HCCMETHOD=2" in proc reg, SAS). $\hat{\beta}_{3,ols}^{[6]}$ is the OLS estimator of $\beta_3^{(6)}$, and $\hat{\sigma}_0^2$ is the overall sample variance of the baseline weight measurement. It follows directly that $\widehat{var}_{aHC}(\hat{\beta}_{1,ols}^{(6)}|\tilde{Y}_{ijt_0})$ is asymptotically unbiased for $var(\hat{\beta}_{1,ols}^{(6)})$ and we can construct a valid test $t = \frac{\hat{\beta}_{1,ols}^{(6)}}{\sqrt{\widehat{var}_{aHC}(\hat{\beta}_{1,ols}^{(6)}|\tilde{Y}_{ijt_0})}}$ for testing $H_o: \tau = 0$ unconditionally.

**Method 7 *ANCOVA* I:** We model the post-treatment weight $Y_{ijt_1}$ using the binary treatment $G$ and baseline measure $Y_{ijt_0}$:

$$Y_{ijt_1} = \beta_0^{(7)} + \beta_1^{(7)}G_{ij} + \beta_2^{(7)}Y_{ijt_0} + e_{ij}^{(7)} \qquad (7)$$

$$e_{i0}^{(7)} \sim N(0, \sigma_{\epsilon_0^{(7)}}^2) \text{ and } \sigma_{\epsilon_0^{(7)}}^2 = (1-\rho_0^2)\sigma_{01}^2 + \left(\beta_3^{(6)}p_1\right)^2\sigma_0^2$$

$$e_{i1}^{(7)} \sim N(0, \sigma_{\epsilon_1^{(7)}}^2) \text{ and } \sigma_{\epsilon_1^{(7)}}^2 = (1-\rho_1^2)\sigma_{11}^2 + \left(\beta_3^{(6)}p_0\right)^2\sigma_0^2$$

,where $\beta_0^{(7)} = \beta_0^{(6)} - \beta_3^{(6)}p_0\mu_0$, and $\beta_1^{(7)} = \tau$. $e_{i0}^{(7)}$ and $e_{i1}^{(7)}$ are random errors in the control and treatment arms. Since $e_{i0}^{(7)}$ and $e_{i1}^{(7)}$ have different variances in general, model (7) is heteroscedastic and the severity of heteroscedasticity is determined by correlation coefficient, variances of post-treatment measurements, and whether the design is balanced.

The OLS estimator $\hat{\beta}_{1,ols}^{(7)}$ is an adjusted mean difference in post-treatment weights controlling for a weighted mean difference of baseline measurements between two arms with equal weights for the treatment and control arms (i.e., $\hat{\beta}_{2,ols}^{(7)}$ for both arms). $\hat{\beta}_{1,ols}^{(7)}$ is unbiased for $\tau$. Similar to *ANCOVA* **II**, OLS model-based inference for *ANCOVA* **I** also assumes a constant residual variance $\sigma_{\epsilon^{(7)}}^2$, which is a weighted average of $\sigma_{\epsilon_0^{(7)}}^2$ and $\sigma_{\epsilon_1^{(7)}}^2$, as follows:

$$\sigma_{\epsilon^{(7)}}^2 = \frac{n_0}{n_0+n_1}\sigma_{\epsilon_0^{(7)}}^2 + \frac{n_1}{n_0+n_1}\sigma_{\epsilon_1^{(7)}}^2.$$

Since $\sigma_{\epsilon^{(7)}}^2$ is unknown, it is estimated by

$$\hat{\sigma}_{\epsilon^{(7)}}^2 = \frac{\sum_{j=0}^1 \sum_{i=1}^{n_j}(y_{ijt_1}-\hat{y}_{ijt_1})^2}{(n_0+n_1-3)},$$

where $\hat{y}_{ijt_1}$ is the predicted value of $y_{ijt_1}$ from fitted model (7). The closed form expressions of $\hat{\beta}_{1,ols}^{(7)}$, the true conditional variance $var(\hat{\beta}_{1,ols}^{(7)}|Y_{ijt_0})$ with two different residual variances, the OLS model-based conditional variance $var_{ols}(\hat{\beta}_{1,ols}^{(7)}|Y_{ijt_0})$ with known common residual variance $\sigma_{\epsilon(7)}^2$, and the OLS model-based variance estimator $\widehat{var}_{ols}(\hat{\beta}_{1,ols}^{(7)}|Y_{ijt_0})$ with the estimator $\hat{\sigma}_{\epsilon(7)}^2$ are given in Table 2. Recall that standard statistical software reports $\widehat{var}_{ols}(\hat{\beta}_{1,ols}^{(7)}|Y_{ijt_0})$. To show model-based standard errors and p-values are valid for unconditional inference, we need to examine: i) whether $\widehat{var}_{ols}(\hat{\beta}_{1,ols}^{(7)}|Y_{ijt_0})$ is unbiased for $var(\hat{\beta}_{1,ols}^{(7)}|Y_{ijt_0})$; ii) whether $var(\hat{\beta}_{1,ols}^{(7)}|Y_{ijt_0})$ is unbiased for $var(\hat{\beta}_{1,ols}^{(7)})$.

First, $\widehat{var}_{ols}(\hat{\beta}_{1,ols}^{(7)}|Y_{ijt_0})$ is unbiased for $var_{ols}(\hat{\beta}_{1,ols}^{(7)}|Y_{ijt_0})$ but the unbiasedness of $\widehat{var}_{ols}(\hat{\beta}_{1,ols}^{(7)}|Y_{ijt_0})$ as an estimator of $var(\hat{\beta}_{1,ols}^{(7)}|Y_{ijt_0})$ depends on the relationship between $var_{ols}(\hat{\beta}_{1,ols}^{(7)}|Y_{ijt_0})$ and $var(\hat{\beta}_{1,ols}^{(7)}|Y_{ijt_0})$. When sample sizes are equal between two arms, we have

$$var_{ols}(\hat{\beta}_{1,ols}^{(7)}|Y_{ijt_0}) \approx var(\hat{\beta}_{1,ols}^{(7)}|Y_{ijt_0}).$$

Thus, $\widehat{var}_{ols}(\hat{\beta}_{1,ols}^{(7)}|Y_{ijt_0})$ is nearly unbiased for $var(\hat{\beta}_{1,ols}^{(7)}|Y_{ijt_0})$ in balanced design[2]. When sample sizes are not equal between two arms,

$$var_{ols}(\hat{\beta}_{1,ols}^{(7)}|Y_{ijt_0}) \neq var(\hat{\beta}_{1,ols}^{(7)}|Y_{ijt_0}),$$

it follows directly that $\widehat{var}_{ols}(\hat{\beta}_{1,ols}^{(7)}|Y_{ijt_0})$ is biased for $var(\hat{\beta}_{1,ols}^{(7)}|Y_{ijt_0})$ due to heteroscedasticity. $\widehat{var}_{ols}(\hat{\beta}_{1,ols}^{(7)}|Y_{ijt_0})$ may over-estimate $var(\hat{\beta}_{1,ols}^{(7)}|Y_{ijt_0})$ when the group with larger residual variance has larger sample size and the group with smaller residual variance has smaller sample size, and otherwise may underestimate $var(\hat{\beta}_{1,ols}^{(7)}|\tilde{Y}_{ijt_0})$[2,3]. *ANCOVA* **I** is robust against heteroscedasticity in balanced design, but not in unbalanced design.

Second, different from *ANCOVA* **II**, $var(\hat{\beta}_{1,ols}^{(7)}|Y_{ijt_0})$ is unbiased for $var(\hat{\beta}_{1,ols}^{(7)})$ because

$$var(\hat{\beta}_{1,ols}^{(7)}) = E(var(\hat{\beta}_{1,ols}^{(7)}|Y_{ijt_0})).$$

Thus, model-based standard errors and associated *p*-values are valid for unconditional inference in balanced design but are biased in unbalanced design. However, this bias can be easily corrected by replacing $\widehat{var}_{ols}(\hat{\beta}_{1,ols}^{(7)}|Y_{ijt_0})$ with an HC variance estimator $\widehat{var}_{HC}(\hat{\beta}_{1,ols}^{(7)}|Y_{ijt_0})$ [16] and corrected *ANCOVA* **I** will provide valid unconditional inference.

***Constrained Repeated Measures heterogeneous variance model ("cRM"):*** We model the baseline and follow up weights $(Y_{ijt_0}, Y_{ijt_1})$ jointly using the binary time point $T_{ij}$, time by treatment interaction $G_{ij} \times T_{ij}$:

$$Y_{ijt} = \gamma_0^{(8)} + \gamma_1^{(8)} T_{ij} + \gamma_2^{(8)} G_{ij} \times T_{ij} + e_{ijt}^{(8)} \quad j = 0,1; i = 1,2,\ldots n_j. \quad (8)$$

$$\begin{pmatrix} e_{i0t_0}^{(8)} \\ e_{i0t_1}^{(8)} \end{pmatrix} \sim N\left(\begin{bmatrix} 0 \\ 0 \end{bmatrix}, \Sigma_0\right) \text{ in the control arm,}$$

$$\begin{pmatrix} e_{i1t_0}^{(8)} \\ e_{i1t_1}^{(8)} \end{pmatrix} \sim N\left(\begin{bmatrix} 0 \\ 0 \end{bmatrix}, \Sigma_1\right) \text{ in the treatment arm,}$$

where $\gamma_0^{(8)} = \mu_{t_0}, \gamma_2^{(8)} = \mu_{0t_1} - \mu_{0t_0}$, and $\gamma_2^{(8)} = \tau$. Noting that subjects in the treatment and control arms have different variance-covariance structures for baseline and follow-up weights, we fit a *cRM* heterogeneous variance GLS model with group specific variance-covariance structure ("repeated/group=" in SAS proc mixed procedure specifies distinct variance-covariance structure for each treatment arm). The formulas of $\hat{\gamma}_{2,gls}^{(8)}$ and $var(\hat{\gamma}_{2,gls}^{(8)})$ are listed in Table 2. The GLS estimator $\hat{\gamma}_{2,gls}^{(8)}$ is asymptotically unbiased for $\gamma_2^{(8)}$. REML is used to derive the empirical or model-based variance estimator $\widehat{var}_{reml}(\hat{\gamma}_{2,gls}^{(8)})$.

*3.3 Methods comparison*

All treatment effect estimators, except ANOVA estimator, are expressed as the mean difference in post-treatment measurements adjusting for the chance imbalance in baseline measurement between two arms in certain ways. Nonetheless, all estimators are unbiased for $\tau$. To compare these competing methods, we evaluate the efficiency of point estimators of treatment effect by comparing their "unconditional" variances. Since the hypothesis testing of no treatment effect is based on dividing the point estimator by its standard error (i.e., variance divided by sample size) and rejecting the null hypothesis when this ratio exceeds a given threshold, the method that produces unbiased point estimate with the smallest unconditional variance is preferred because standard error in the dominator of statistical test determines the statistical power.

*3.3.1 When study population is homogeneous*

*ANCOVA* **I** is a more efficient alternative to *ANOVA* because $var(\hat{\beta}_{1,ols}^{(2)}) \leq var(\hat{\beta}_{1,ols}^{(1)})$ (Table 1). This advantage of ANCOVA over ANOVA can also be observed from the fact that residual error variance of *ANCOVA* **I** is less than residual error variance of *ANOVA* (i.e., $(1-\rho^2)\sigma_1^2 \leq \sigma_1^2$). When the correlation coefficient $\rho$ becomes larger, the *ANCOVA* **I** estimator has smaller variance. Since $Y_{ijt_1}$ and $Y_{ijt_0}$ are highly correlated in general, the inclusion of $Y_{ijt_0}$ in *ANCOVA* **I** explains away some variability in $Y_{ijt_1}$ and thus reduces residual variance and yields a more efficient estimator of treatment effect than *ANOVA*.

*ANOVA-Change* and *RM* have exactly same point estimators of $\tau$ and have the same variances (Table 1). To compare *ANOVA-Change* or *RM* with *ANOVA*, we can derive the difference between their variances as follows:

$$\Delta = \sigma_0(1 - 2\rho\sigma_1).$$

When $\rho < \frac{1}{2\sigma_1}$, $\Delta > 0$ and *ANOVA* outperforms *ANOVA-Change* and *RM* because *ANOVA* estimator has smaller variance. When $\rho > \frac{1}{2\sigma_1}$, $\Delta < 0$ and *ANOVA* underperforms the other two methods.

It can be shown that the difference between ("unconditional") variances of the *ANCOVA* I or *cRM* estimators and those of *ANOVA-Change* or *RM* estimators are always nonnegative:

$$\Delta = (\sigma_1^2 + \sigma_0^2 - 2\rho\sigma_0\sigma_1) - (1-\rho^2)\sigma_1^2 = (\sigma_0 - \rho\sigma_1)^2 \geq 0$$

Thus, *ANOVA-Change* or *RM* is less efficient than either *ANCOVA* I or *cRM* because their estimators have larger variances. Intuitively *ANCOVA* I or *cRM* assumes two arms have the same mean baseline weights in a randomized study but *ANOVA-Change* or *RM* assumes there is a baseline difference and needs to estimate one extra parameter.

As shown in Table 1, the *ANCOVA* I and *cRM* estimators of $\tau$ are equivalent because $\beta_{1,ols}^{(2)} = \frac{\rho\sigma_0\sigma_1}{\sigma_0^2}$. However, *ANCOVA* I plugs in the OLS estimators $\hat{\beta}_{1,ols}^{(2)}$, whereas *cRM* plugs in the REML estimators of variance and covariance parameters. The numerical difference between $\hat{\beta}_{1,ols}^{(2)}$ and $\hat{\gamma}_{3,gls}^{(4)}$ becomes negligible as sample size increases. Because of this equivalence between $\hat{\beta}_{1,ols}^{(2)}$ and $\hat{\gamma}_{3,gls}^{(4)}$, $var(\hat{\beta}_{1,ols}^{(2)})$ and $var(\hat{\gamma}_{3,gls}^{(4)})$ are equal[2]. As discussed in section (3.1), *ANCOVA* I is a conditional model assuming fixed baseline covariates. Even though model-based variance estimates are conditional, the usual model-based conditional inference is still valid for unconditional hypothesis testing and *ANCOVA* I performs comparably to *cRM*.

3.3.2 When study population is heterogeneous

A heterogeneous study population justifies the inclusion of a treatment by baseline weight interaction term. Thus, *ANCOVA* II is the correctly specified model, whereas *ANCOVA* I is a mis-specified model.

In this case, the treatment effect is not assumed to be constant across different values of baseline weight, known as conditional treatment effect. The ("marginal") treatment effect $\tau$ is simply the average of conditional treatment effect over the distribution of baseline weight and measures an overall treatment effect. As shown in sections (3.2), both ANCOVA models can be used to estimate $\tau$ even though *ANCOVA* **I** is mis-specified. Then, what is the advantage of using the more complex interaction model over a main effect model? It turns out the *ANCOVA* **II** estimator $\hat{\beta}_{1,ols}^{(6)}$ is more efficient than the *ANCOVA* **I** estimator $\hat{\beta}_{1,ols}^{(7)}$ because $var(\hat{\beta}_{1,ols}^{(6)}) \leq var(\hat{\beta}_{1,ols}^{(7)})$ [3]. Only when design is balanced, $var(\hat{\beta}_{1,ols}^{(6)}) = var(\hat{\beta}_{1,ols}^{(7)})$ and the two ANCOVA models perform comparably. Note that the OLS model-based variance estimates for *ANCOVA* **I** and **II** are biased for the corresponding unconditional variances but HC-variance estimators provide simple fixes.

The *ANCOVA* **II** and *cRM* estimators of $\tau$ are equivalent because $\beta_2^{(6)} + \beta_3^{(6)} = \frac{\rho_0 \sigma_0 \sigma_{01}}{\sigma_0^2}$ and $\beta_2^{(6)} = \frac{\rho_1 \sigma_0 \sigma_{11}}{\sigma_0^2}$ (Table 2). Two methods only differ in the way two estimators are estimated. *ANCOVA* **II** plugs in the OLS estimators $\hat{\beta}_{2,ols}^{(6)}$ and $\hat{\beta}_{3,ols}^{(6)}$, whereas *cRM* plugs in the REML estimators of variance and covariance parameters. The numerical difference between the *ANCOVA* **II** and *cRM* estimators gets smaller as sample size increases. As discussed in section (3.2), standard statistical software such as SAS does not output unconditional variance for *ANCOVA* **II** directly but the usual OLS model-based standard errors and *p*-values are biased for unconditional inference in heterogeneous scenario. Adjusted HC-variance estimator fixes this bias. Corrected *ANCOVA* **II** provides valid unconditional inference and performs comparably to *cRM.* Another alternative approach to estimate variances of the *ANCOVA* **I and II** estimators is to use bootstrap method[17].

## 4. Numerical illustration

We simulated three weight loss trial data sets based on a published study for three scenarios: homogeneous data, heterogeneous data with balanced and unbalanced design as follows[18]:

1) The baseline weights for the control and treatment arms are generated from normal distribution with mean 88 kg and standard deviation 14 kg. Weights at 6 month after treatment for the control arm have mean 86 kg and standard deviation 15 kg. This gives a ~2.3% change from baseline. Mean and standard deviation for weights at 6 month in the treatment arm are 83 kg and 15 kg, respectively; This corresponds to a 5.7% change.

2) In homogeneous data, the correlation coefficient between baseline and follow-up weights is 0.9. 180 subjects were assigned to the treatment and control arms equally. In heterogeneous data, the correlation coefficient between the baseline and follow-up weights in the control arm is 0.9 and 0.7 in the treatment arm. Sample sizes are ($n_0 = 90, n_1 = 90$) for balanced design and ($n_0 = 60, n_1 = 120$) for unbalanced design. We analyzed the data examples using the methods outlined in section (3). The statistical results were reported in Table 3 (SAS programs are provided in the Appendix).

In the first data example, *ANOVA* produced the largest standard error and the largest *p*-value. *ANOVA-Change* and *RM* both outperformed *ANOVA* with much smaller standard errors and *p*-values. *ANCOVA* **I** and *cRM* outperformed *ANOVA-Change* and *RM* with smaller standard errors and *p*-values. Although *ANCOVA* **I** and *cRM* are equivalent when sample size is large, there are still small differences between the two in finite sample.

For the second data example with a balanced design, Figure (2.a) shows that there is a strong baseline weight by treatment interaction. Both *ANCOVA* **I** and **II** have heteroscedastic errors by treatment arm (Figure (2.b) and (2.c)). As shown in table 2, OLS model-based standard error of *ANCOVA* **I** is very similar to its HC and bootstrap standard errors. Thus, heteroscedasticity does not bias model-based standard error of *ANCOVA* **I**. Although *ANCOVA* **II** is robust against heteroscedasticity in balanced design, OLS model-based standard error of *ANCOVA* **II** ($s.e$=1.333) is still not correct because OLS fails to consider the variability of estimating the overall mean baseline weight. Adjusted HC standard error for *ANCOVA* **II** is 1.402, which is closer to model-based and HC standard errors of *ANCOVA* **I**. Bootstrapping standard errors for *ANCOVA* **I** and **II** are close to their HC or adjusted HC standard errors. *cRM* estimate and its standard error are close to those from *ANCOVA* **I** and **II**.

For the third example with an unbalanced design, Figure (3.d) also reveals a baseline weight by treatment interaction. Both ANCOVA models have heteroscedastic errors by treatment arm (Figure (3.e) and (3.f)). Model-based standard errors of both ANCOVA models are not valid. Model-based standard errors are larger than HC standard errors and thus overestimated the true conditional variances. Compared with *ANCOVA* **I**, *ANCOVA* **II** has a smaller HC standard error (also smaller *p*-value) and thus is slightly more efficient. Adjusted HC standard error for *ANCOVA* **II** is very close to model-based standard error for *cRM*. Bootstrapping standard errors for *ANCOVA* **I** and **II** are very close to their HC or adjusted HC standard errors.

## *5. Discussion*

In this study we compare the efficiency of six unbiased methods analyzing pre-post design. We found ANCOVA and *cRM* are the equally most efficient methods compared with other alternatives in homogeneous and heterogeneous scenarios.

The majority of previous studies only examined homogeneous study population. In this setting, *ANOVA* is one of the least efficient approaches for analyzing pre-post designs because it does not utilize any baseline information. *ANOVA-Change* and *RM* incorporate baseline measurement as part of outcome, whereas *ANCOVA* **I** adjusts baseline measurement as regression adjustment. *ANCOVA* **I** outperforms *ANOVA-Change* and *RM* because *ANCOVA* **I** utilizes the assumption of the balanced baseline measurement between two arms in a randomized study. Thus, change score is a less efficient way to utilize baseline information than adjusting baseline variables as covariates. Because in randomized trials we seldom can control the values of baseline measurement, the assumption that baseline measurement is fixed required by OLS casts doubt on the validity of ANCOVA for hypothesis testing[5,10]. Crager proved *ANCOVA* **I** is valid for unconditional inference in homogeneous scenario[5]. This conclusion can be simply attributed to that the conditional variance of the *ANCOVA* **I** estimator is an unbiased estimate for its unconditional variance[2].

A few studies investigated further a heterogeneous scenario[2,3,9,10,19]. Although the heterogeneity justifies the inclusion of the baseline measurement by treatment interaction term, *ANCOVA* **I** and **II** are both unbiased. Yang and Tsiatis showed the *ANCOVA* **II** has smaller unconditional variance than that of the *ANCOVA* **I** estimator unless in balanced design[20]. However, the OLS model-based variances of *ANCOVA* **I** and **II** estimators, output by standard statistical software, are conditional variances, not unconditional variances. The OLS model-based standard errors and associated *p*-values for *ANCOVA* **I** and **II** are generally questionable for unconditional inference[2,3,9,19], particularly when the design is unbalanced. With corrected HC variance estimators, both models provide valid unconditional inference. Choosing between *ANCOVA* **I** and **II** then becomes an evaluation of a trade-off between simplicity and some gains in efficiency.

In homogenous setting, *cRM* was suggested as a superior choice to *ANCOVA* **I** because unconditional variance of the *cRM* estimator is smaller than conditional variance of the *ANCOVA* **I** estimator[21]. Kenward et al. pointed out that such direct comparison between conditional and unconditional variances is not meaningful. Both estimators are equivalent and *cRM* based on REML and Kenward-roger adjustment performed almost identically to *ANCOVA* **I** in finite samples[15]. In heterogeneous scenario, *cRM* is comparable to *ANCOVA* **II**[2]. In presence of missing data, applied researchers often prefer *cRM* over ANCOVA because it can utilize all observed data but ANCOVA uses only complete cases. However, imputation methods which utilize the strong pre-post correlation, such as weighting and regression imputation, can improve the statistical power for ANCOVA without biasing estimates, making it comparable to *cRM*[15].

Furthermore, ANCOVA has several advantages over *cRM*: first, outcome should only be the variable that can be influenced by treatment. Baseline measurement is certainly not an outcome by this definition. It is conceptually more appropriate to adjust baseline measurement as covariate, not model it as outcome[4]; Second, it is very convenient to include other baseline variables in a regression model for more efficient estimates of treatment effect. Third, it is easy to adjust for other patterns of heteroscedastic errors in an OLS regression. For example, we may expect larger variability in post-treatment weight associated with larger baseline weight. *cRM* cannot handle this more complex type of heteroscedasticity easily. HC-variance estimators for ANCOVA are simple fixes and readily implemented in statistical statistical software.

Figure 1. Hypothetical two arm pre-post weight loss randomized trial

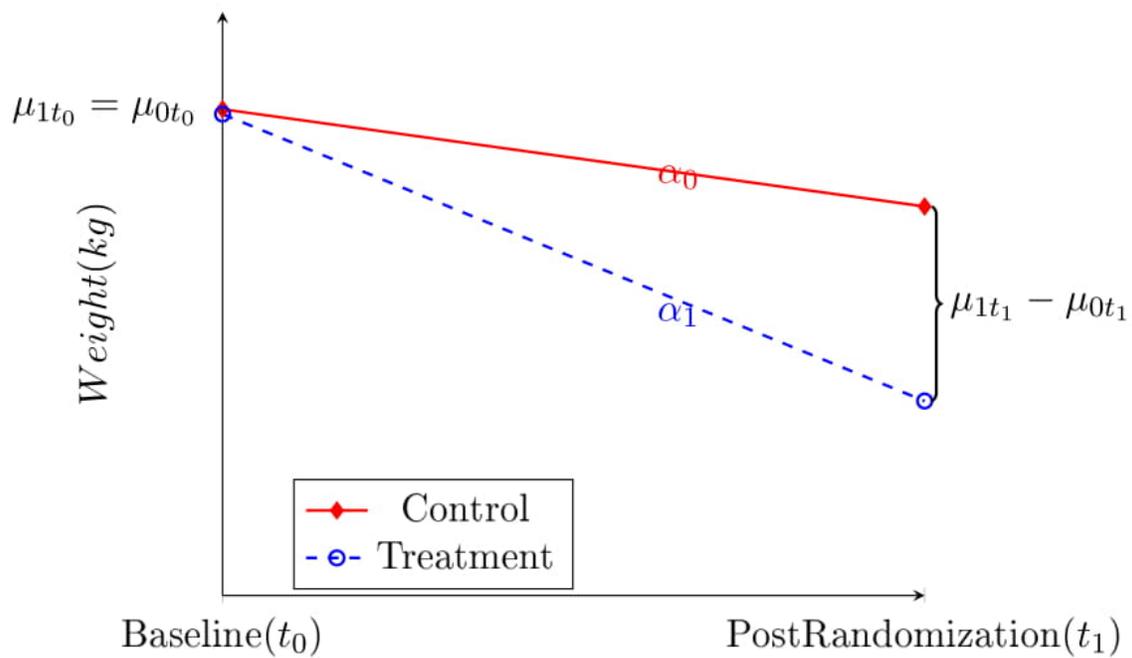

Figure 2. Diagnosis plots of ANCOVA main and interaction models in heterogeneous scenario.

*a*)Scatter plot of baseline and follow-up weights in balanced design. Black and red solid dots are data points in the treatment and control arms. Black and red solid lines are regression slopes of baseline weight against follow-up weight in the treatment and control arms. *b*) Boxplot of residuals from the treatment and control arms from *ANCOVA* **I** model in balanced design; *c*) Boxplot of residuals from the treatment and control arms from *ANCOVA* **II** model in balanced design; *d*) Scatter plot of baseline and follow-up weights in unbalanced design. Black and red solid dots are data points in the treatment and control arms. Black and red solid lines are regression slopes of baseline weight against follow-up weight in the treatment and control arms. *e*) Boxplot of residuals from the treatment and control arms from *ANCOVA* **I** model in unbalanced design; *f*) Boxplot of residuals from the treatment and control arms from *ANCOVA* **II** model in unbalanced design;

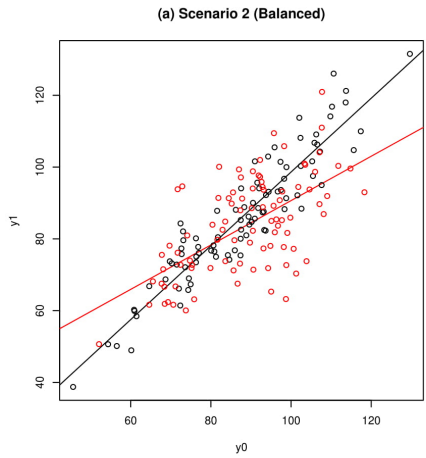 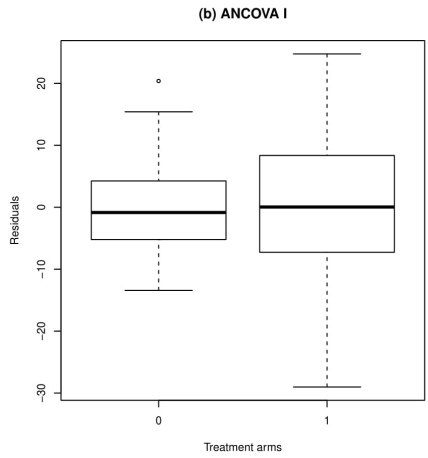 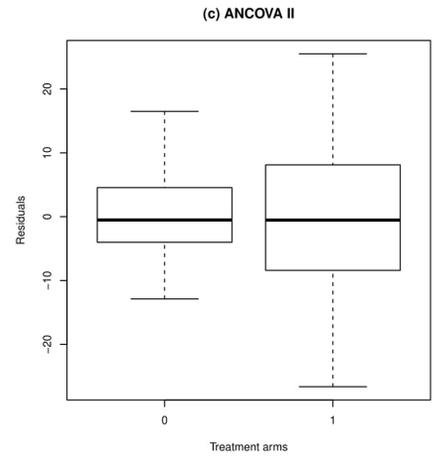 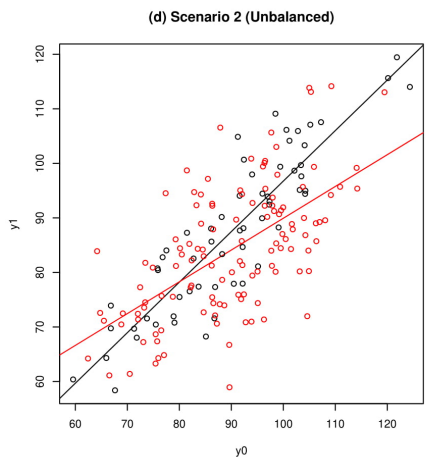 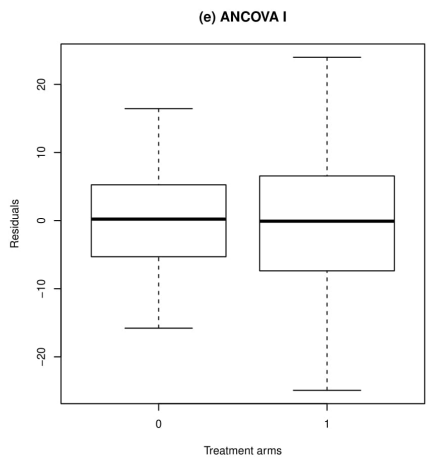 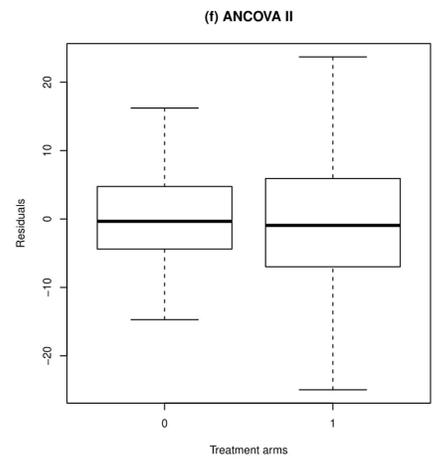

**Table 1 Estimators of treatment effect and variance estimators in a homogeneous study population**

| Model | Estimator of Treatment Effect ($\tau$) | Type* | True Variance of Estimator | OLS Model Based Variance Estimator |
|---|---|---|---|---|
| ANOVA-Post | $\hat{\beta}_{1,ols}^{(1)} = \bar{y}_{.1t_1} - \bar{y}_{.0t_1}$ | U | $var(\hat{\beta}_{1,ols}^{(1)}) = \frac{\sigma_1^2}{n_0} + \frac{\sigma_1^2}{n_1}$ | $\widehat{var}_{ols}(\hat{\beta}_{1,ols}^{(1)}) = \frac{\hat{\sigma}_1^2}{\sum_{j=0}^{1}\sum_{i=1}^{n_j}(G_{ij} - G_{..})^2}$ <br> $\hat{\sigma}_1^2 = \frac{\sum_{j=0}^{1}\sum_{i=1}^{n_j}(y_{ijt_1} - \hat{y}_{ijt_1})^2}{(n_0 + n_1 - 2)}$ |
| ANCOVA-Post I | $\hat{\beta}_{1,ols}^{(2)} = (\bar{y}_{.1t_1} - \bar{y}_{.0t_1}) - \hat{\beta}_{2,ols}^{(2)}(\bar{y}_{.1t_0} - \bar{y}_{.0t_0})$ | C | $var(\hat{\beta}_{1,ols}^{(2)}|Y_{ijt_0}) = \left(\frac{1}{n_0} + \frac{1}{n_1} + \frac{(\bar{y}_{.1t_0}-\bar{y}_{.0t_0})^2}{\sum_{j=0}^{1}\sum_{i=1}^{n_j}(y_{ijt_0}-\bar{y}_{.jt_0})^2}\right)\sigma_{\epsilon^{(2)}}^2$, <br><br> $\sigma_{\epsilon^{(2)}}^2 = (1-\rho^2)\sigma_1^2$ | $\widehat{var}_{ols}(\hat{\beta}_{1,ols}^{(2)}|Y_{ijt_0}) = \left(\frac{1}{n_0} + \frac{1}{n_1} + \frac{(\bar{y}_{.1t_0}-\bar{y}_{.0t_0})^2}{\sum_{j=0}^{1}\sum_{i=1}^{n_j}(y_{ijt_0}-\bar{y}_{.jt_0})^2}\right)\hat{\sigma}_{e_{ij}^{(2)}}^2$, <br><br> $\hat{\sigma}_{e_{ij}^{(2)}}^2 = \frac{\sum_{j=0}^{1}\sum_{i=1}^{n_j}(y_{ijt_1} - \hat{y}_{ijt_1})^2}{(n_0 + n_1 - 4)}$ |
| | | U | $var(\hat{\beta}_{1,ols}^{(2)}) = \left(\frac{1}{n_0} + \frac{1}{n_1}\right)(1-\rho^2)\sigma_1^2$ | |
| RM | $\hat{\gamma}_{3,gls}^{(3)} = (\bar{y}_{.1t_1} - \bar{y}_{.1t_0}) - (\bar{y}_{.0t_1} - \bar{y}_{.0t_0})$ | U | $var(\hat{\gamma}_{3,gls}^{(3)}) = (\frac{1}{n_0} + \frac{1}{n_1})(\sigma_1^2 + \sigma_0^2 - 2\rho\sigma_0\sigma_1)$ | |
| cRM | $\hat{\gamma}_{3,gls}^{(4)} = (\bar{y}_{.1t_1} - \bar{y}_{.0t_1}) - \frac{\rho\sigma_0\sigma_1}{\sigma_0^2}(\bar{y}_{.1t_0} - \bar{y}_{.0t_0})$ | U | $var(\hat{\gamma}_{3,gls}^{(4)}) = \left(\frac{1}{n_0} + \frac{1}{n_1}\right)(1-\rho^2)\sigma_1^2$ | |
| ANOVA-Change | $\hat{\beta}_{1,ols}^{(5)} = (\bar{y}_{.1t_1} - \bar{y}_{.1t_0}) - (\bar{y}_{.0t_1} - \bar{y}_{.0t_0})$ | U | $var(\hat{\beta}_{1,ols}^{(5)}) = (\frac{1}{n_0} + \frac{1}{n_1})(\sigma_1^2 + \sigma_0^2 - 2\rho\sigma_0\sigma_1)$ | $\widehat{var}_{ols}(\hat{\beta}_{1,ols}^{(5)}) = \frac{\hat{\sigma}_{\epsilon^{(5)}}^2}{\sum_{j=0}^{1}\sum_{i=1}^{n_j}(G_{ij} - G_{..})^2}$, <br> $\hat{\sigma}_{\epsilon^{(5)}}^2 = \frac{\sum_{j=0}^{1}\sum_{i=1}^{n_j}(\Delta_{ij} - \hat{\Delta}_{ij}^{(5)})^2}{(n_0 + n_1 - 2)}$ |

* U- unconditional variance; C- conditional variance

Table 2: **Estimators of treatment effect and variance estimators in a heterogeneous study population**

| Model | Estimator of Treatment Effect ($\tau$) | Type | True Variance of Estimator | Variance Estimator from OLS Model |
|---|---|---|---|---|
| ANCOVA-Post II | $\hat{\beta}_{1,ols}^{(6)} = (\bar{y}_{.1t_1} - (\hat{\beta}_{2,ols}^{(6)} + \hat{\beta}_{3,ols}^{(6)})\bar{\tilde{y}}_{.1t_0}) - (\bar{y}_{.0t_0} - \hat{\beta}_{2,ols}^{(6)}\bar{\tilde{y}}_{.0t_0})$ | C | $var(\hat{\beta}_{1,ols}^{(6)}|\tilde{Y}_{ijt_0}) = \left(\frac{1}{n_0} + \frac{\bar{\tilde{y}}_{.0t_0}^2}{\sum_{i=1}^{n_0}(\tilde{y}_{i0t_0} - \bar{\tilde{y}}_{.0t_0})^2}\right)\sigma_{\epsilon_0^{(6)}}^2$ $+ \left(\frac{1}{n_1} + \frac{\bar{\tilde{y}}_{.1t_0}^2}{\sum_{i=1}^{n_0}(\tilde{y}_{i1t_0} - \bar{\tilde{y}}_{.1t_0})^2}\right)\sigma_{\epsilon_1^{(6)}}^2$ $\sigma_{\epsilon_0^{(6)}}^2 = (1-\rho_0^2)\sigma_{01}^2, \quad \sigma_{\epsilon_1^{(6)}}^2 = (1-\rho_1^2)\sigma_{11}^2$ | $\widehat{var}_{ols}(\hat{\beta}_{1,ols}^{(6)}|\tilde{Y}_{ijt_0})$ $= \left(\frac{1}{n_0} + \frac{1}{n_1} + \frac{\bar{\tilde{y}}_{.0t_0}^2}{\sum_{i=1}^{n_0}(\tilde{y}_{i0t_0} - \bar{\tilde{y}}_{.0t_0})^2}\right.$ $\left.+ \frac{\bar{\tilde{y}}_{.1t_0}^2}{\sum_{i=1}^{n_0}(\tilde{y}_{i1t_0} - \bar{\tilde{y}}_{.1t_0})^2}\right)\hat{\sigma}_{\epsilon^{(6)}}^2$ $\hat{\sigma}_{\epsilon^{(6)}}^2 = \frac{\sum_{j=0}^{1}\sum_{i=1}^{n_j}(y_{ijt_1} - \hat{y}_{ijt_1})^2}{(n_0 + n_1 - 5)}$ |
| | | U | $var(\hat{\beta}_{1,ols}^{(6)}) = \frac{1}{n_0}(1-\rho_0^2)\sigma_{01}^2 + \frac{1}{n_1}(1-\rho_1^2)\sigma_{11}^2$ $+ \left(\rho_1\frac{\sigma_{11}}{\sigma_0} - \rho_0\frac{\sigma_{01}}{\sigma_0}\right)^2\frac{\sigma_0^2}{n_0+n_1}$ | |
| ANCOVA-Post I | $\hat{\beta}_{1,ols}^{(7)} = (\bar{y}_{.1t_1} - \bar{y}_{.0t_1}) - \hat{\beta}_{2,ols}^{(7)}(\bar{y}_{.1t_0} - \bar{y}_{.0t_0})$ | C | $var(\hat{\beta}_{1,ols}^{(7)}|Y_{ijt_0}) = \left(\frac{1}{n_0} + \frac{\sum_{i=1}^{n_0}(y_{i1t_0}-\bar{y}_{.0t_0})^2(\bar{y}_{.1t_0}-\bar{y}_{.0t_0})}{\sum_{j=0}^{1}\sum_{i=1}^{n_j}(y_{ijt_0}-\bar{y}_{.jt_0})^2}\right)\sigma_{\epsilon_0^{(7)}}^2$ $+ \left(\frac{1}{n_1}\right.$ $\left.+ \frac{\sum_{i=1}^{n_1}(y_{i1t_0}-\bar{y}_{.1t_0})^2(\bar{y}_{.1t_0}-\bar{y}_{.0t_0})}{\sum_{j=0}^{1}\sum_{i=1}^{n_j}(y_{i1t_0}-\bar{\tilde{y}}_{.1t_0})^2}\right)\sigma_{\epsilon_1^{(7)}}^2$ $\sigma_{\epsilon_0^{(7)}}^2 = (1-\rho_0^2)\sigma_{01}^2, \quad \sigma_{\epsilon_1^{(7)}}^2 = (1-\rho_1^2)\sigma_{11}^2$ | $\widehat{var}_{ols}(\hat{\beta}_{1,ols}^{(7)}|Y_{ijt_0})$ $= \left(\frac{1}{n_0} + \frac{1}{n_1} + \frac{\sum_{i=1}^{n_0}(y_{i1t_0}-\bar{y}_{.0t_0})^2(\bar{y}_{.1t_0}-\bar{y}_{.0t_0})}{\sum_{j=0}^{1}\sum_{i=1}^{n_j}(y_{ijt_0}-\bar{y}_{.jt_0})^2}\right.$ $\left.+ \frac{\sum_{i=1}^{n_1}(y_{i1t_0}-\bar{y}_{.1t_0})^2(\bar{y}_{.1t_0}-\bar{y}_{.0t_0})}{\sum_{j=0}^{1}\sum_{i=1}^{n_j}(y_{i1t_0}-\bar{\tilde{y}}_{.1t_0})^2}\right)\hat{\sigma}_{\epsilon^{(7)}}^2$ $\hat{\sigma}_{\epsilon^{(7)}}^2 = \frac{\sum_{j=0}^{1}\sum_{i=1}^{n_j}(y_{ijt_1} - \hat{y}_{ijt_1})^2}{(n_0 + n_1 - 4)}$ |
| | | U | $var(\hat{\beta}_{1,ols}^{(7)}) = \frac{1}{n_0}[(1-\rho_0^2)\sigma_{01}^2 + ((\rho_1\frac{\sigma_{11}}{\sigma_0} - \rho_0\frac{\sigma_{01}}{\sigma_0})p_1)^2\sigma_0^2 +$ $\frac{1}{n_1}[(1-\rho_1^2)\sigma_{11}^2 + ((\rho_1\frac{\sigma_{11}}{\sigma_0} - \rho_0\frac{\sigma_{01}}{\sigma_0})p_0)^2\sigma_0^2]$ | |
| cRM | $\hat{\gamma}_{3,gls}^{(4)} = (\bar{y}_{.1t_1} - \bar{y}_{.0t_1}) - (\frac{\rho_0\sigma_0\sigma_{01}}{\sigma_0^2}(\bar{y}_{.1t_0} - \bar{y}_{..t_0}) - \frac{\rho_1\sigma_0\sigma_{11}}{\sigma_0^2}(\bar{y}_{.1t_0} - \bar{y}_{..t_0}))$ | U | $var(\hat{\gamma}_{3,gls}^{(4)}) = \frac{1}{n_0}[(1-\rho_0^2)\sigma_{01}^2 + ((\rho_1\frac{\sigma_{11}}{\sigma_0} - \rho_0\frac{\sigma_{01}}{\sigma_0})p_1)^2\sigma_0^2 +$ $\frac{1}{n_1}[(1-\rho_1^2)\sigma_{11}^2 + ((\rho_1\frac{\sigma_{11}}{\sigma_0} - \rho_0\frac{\sigma_{01}}{\sigma_0})p_0)^2\sigma_0^2]$ | |

Table 3- Statistical analysis of the three simulated data examples

| Scenario | Method | Estimate | Standard error | p-value |
|---|---|---|---|---|
| Homogeneous | ANOVA | -3.089 | 2.106 | 0.144 |
| | ANCOVA I | -2.422 | 0.955 | 0.0121 |
| | ANOVA-Change | -2.354 | 0.971 | 0.0163 |
| | RM | -2.354 | 0.971 | 0.0163 |
| | cRM | -2.434 | 0.944 | 0.0108 |
| Heterogeneous ($n_0$=90, $n_1 = 90$) | ANCOVA I | -3.203 | 1.403[1] | 0.0235 |
| | | | 1.397[2] | 0.0231 |
| | | | 1.400[4] | n/a |
| | ANCOVA II | -3.165 | 1.333[1] | 0.0187 |
| | | | 1.402[3] | 0.0252 |
| | | | 1.397[4] | n/a |
| | cRM | -3.203 | 1.405 | 0.0241 |
| Heterogeneous ($n_0$=60, $n_1 = 120$) | ANCOVA I | -3.416 | 1.415[1] | 0.0167 |
| | | | 1.279[2] | 0.0083 |
| | | | 1.281[4] | n/a |
| | ANCOVA II | -3.399 | 1.376[1] | 0.0145 |
| | | | 1.258[3] | 0.0076 |
| | | | 1.260[4] | n/a |
| | cRM | -3.396 | 1.262 | 0.0078 |

[1] OLS regression model-based standard error

[2] HC standard error for ANCOVA I (main effect) model

[3] Modified HC standard error for ANCOVA II (interaction) model

[4] Bootstrapping standard error (n=5000)

Appendix

SAS programs for ANCOVA I, ANCOVA II, and cRM in a heterogeneous scenario

*****compute overall mean of baseline measurement****;

**proc means** data=data3 mean std stderr;

   var y0 ;

   output out=meandata mean=y0bar stderr=y0err;

  **run**;

***create mean centered baseline variable y0cen***;

**data** data3_1;

   set data3;

   y0cen=y0-**90.0817872**;

   int_y0cen=trt*y0cen;

  **run**;

***ANCOVA I including main effect of treament and baseline variable y0***;

***use HHMETHOD=2 to compute HC2 variance***;

**proc reg** data=data3_1;

    model y1=trt y0/white HCCMETHOD=**2**;

    **run**;

***ANCOVA II including treatment, y0cen, and the interaction term***;

***use HHMETHOD=2 to compute HC2 variance***;

**proc reg** data=data3_1;

   ods output ParameterEstimates=perobust;

   model y1=trt y0cen int_y0cen/white HCCMETHOD=**2**;

  **run**;

```sas
data perobust2;
    set perobust;
    keep variable estimate stderr tvalue probt hcstderr;
    run;

proc sql noprint;
    select y0bar into: y0bar
        from meandata;
    select y0err into: y0err
        from meandata;

%put &y0bar &y0err;

proc sql noprint;
    select estimate into: beta0
        from perobust2
        where variable="Intercept";
    select estimate into: beta1
        from perobust2
        where variable="trt";
    select estimate into: beta2
        from perobust2
        where variable="y0cen";
    select estimate into: beta3
        from perobust2
        where variable="int_y0cen";

data perobust3;
    set perobust2;
        * if variable="Intercept" then modified_hcstderr=sqrt(hcstderr**2+((&beta2)**2)*((&y0err)**2));
```

```sas
        if variable="trt"     then modified_hcstderr=sqrt(hcstderr**2+((&beta3)**2)*((&y0err)**2));
      * if variable="y0cen"    then modified_hcstderr=hcstderr;
      * if variable="int_y0cen" then modified_hcstderr=hcstderr;
        corrZ=estimate/modified_hcstderr;
        pvalue=2*(1-probnorm(abs(corrZ)));
        if variable in ("Intercept","trt")  then pvalue2=2*(1-probt(abs(corrz),176));
       *if variable in ("y0cen","int_y0cen") then pvalue2=2*(1-probt(abs(corrz),299));
        if  variable="trt" ;
        keep variable estimate stderr tvalue probt hcstderr modified_hcstderr corrz pvalue pvalue2;
     run;

data mixeddata3;
    set data3;
    do; y=y1;time=1;time2=time;output;end;
    do; y=y0;time=0;time2=time;output;end;
    run;

title "constrained repeated measures (cRM) with heterogeneous variance";
***analysis 6- conditional longitudinal analysis (fix arm term to be 0)***;
proc mixed data=mixeddata3 method=reml order=data cl=wald;
    class time subj;
    model y=time2 trt*time2/ddfm=kenwardroger solution alpha=0.05;
    repeated time/subject=subj type=un group=trt rcorr;
    run;
```